\documentclass[aps,prd,twocolumn,nofootinbib,superscriptaddress]{revtex4-1} 

\usepackage{amsmath,amssymb}
\usepackage{graphicx}

\begin{document}
\title{Nonflat time-variable dark energy cosmology}

\author{Anatoly Pavlov}
\email{pavlov@phys.ksu.edu}
\affiliation{Department of Physics, Kansas State University, 116 Cardwell Hall, Manhattan, Kansas 66506, USA}

\author{Shawn Westmoreland}
\email{westmore@phys.ksu.edu}
\affiliation{Department of Physics, Kansas State University, 116 Cardwell Hall, Manhattan, Kansas 66506, USA}

\author{Khaled Saaidi}
\email{khaledsaeidi@gmail.com}
\affiliation{Department of Physics, Faculty of Science, University of Kurdistan, Sanandaj 66177-15177, Iran}
\affiliation{Department of Physics, Kansas State University, 116 Cardwell Hall, Manhattan, Kansas 66506, USA}

\author{Bharat Ratra}
\email{ratra@phys.ksu.edu}
\affiliation{Department of Physics, Kansas State University, 116 Cardwell Hall, Manhattan, Kansas 66506, USA}
\date{\today ~~~ KSUPT - 13/3}

\begin{abstract}

We generalize the time-variable dark energy scalar field $\Phi$ model ($\Phi$CDM) to nonflat space. We show that even in the space-curvature-dominated epoch the scalar field solution is a time-dependent fixed point or attractor, with scalar field energy density that grows relative to the energy density in spatial curvature. This is the first example of a physically consistent and complete model of dynamical dark energy in a nonflat geometry.
\end{abstract}

\maketitle
\section{Introduction}
\label{Intro}

When measurements of the cosmic microwave background (CMB) anisotropy are examined in the context of the current standard model of cosmology, the $\Lambda$CDM model\footnote{In this model \cite{Peebles}, the current cosmological energy budget is dominated by a cosmological constant $\Lambda$, with non-relativistic cold dark matter (CDM) being the next largest contributor. For some time now most observations have been reasonably consistent with the predictions of the spatially-flat $\Lambda$CDM model; see for example \cite{MultRefs_1}. Note that there are tentative observational indications that the standard CDM structure formation model, assumed in the $\Lambda$CDM cosmological model, might need to be improved upon \cite{Weinberg}.}, they indicate that the cosmological spatial hypersurfaces are close to flat \cite{Ade}. On the other hand, under the assumption of flat spatial geometry the data favor time-independent dark energy (DE). However, it has been known for a while now that if a spatially curved time-variable DE model is used to analyze the CMB anisotropy measurements there is a degeneracy between spatial curvature and the parameter that governs the DE time variability, and this results in significantly weaker constraints on both parameters compared to the cases when only either non-zero spatial curvature or DE time variability is assumed \cite{MultRefs_2}.

Most of these analyses are based on the XCDM parametrization or generalizations thereof. In the XCDM parametrization time-evolving DE is taken to be an $X$-fluid with equation of state $p_X = w_X \rho_X$ where $\rho_X$ and $p_X$ are the $X$ fluid energy density and pressure and the equation of state parameter $w_X < -1/3$ is a constant. This is an incomplete model of time-variable DE since, unless extended, it cannot consistently describe the evolution of spatial inhomogeneities \cite{RatraPRD1991}.

The $\Phi$CDM model \cite{Peebles&Ratra, Ratra&PeeblesPRD} is the simplest consistent model of time-variable DE. In this model a scalar field $\Phi$ with potential energy density $V(\Phi)$ is the DE; $V(\Phi) \propto \Phi^{-\alpha}$, where constant $\alpha > 0$, is a widely used example\footnote{See Ref. \cite{JMartin2008} for more general examples.}. The original $\Phi$CDM model assumed flat spatial sections. In this paper we develop the curved space extension of the $\Phi$CDM model. Related models have been previously considered, see Ref.\ \cite{MultRefs_3}. However, as far as we are aware, we are the first to establish that the scalar field solution in the curvature-dominated epoch is a time-dependent fixed point or attractor, and that in the curvature-dominated epoch the scalar field energy density grows relative to that of space curvature, generalizing the results of \cite{Peebles&Ratra, Ratra&PeeblesPRD} to curved space.

Our paper is organized as follows. In the next section we describe the curved-space $\Phi$CDM model we study. In this section and in the Appendix we show that this model has a time-dependent fixed point scalar field solution in the curvature-dominated epoch. In Sec.\ \ref{CosTests} we compute some observable cosmological-test predictions for this model as a function of the three cosmological parameters of the model. Then we discuss these results by comparing those for flat and nonflat geometries as well as for open and closed geometries. In the final section we provide conclusions.

\section{The Model}
\label{Model}
The original $\Phi$CDM model of \cite{Peebles&Ratra} was designed to describe the late-time consequences of an inflationary scalar field $\Phi$ model in which the scalar field potential energy density $V(\Phi)$ has an inverse power-law tail at large $\Phi$. This form of $V(\Phi)$ was chosen because it provides a self-consistent phenomenological description of DE whose density decreases as the Universe expands, but decreases less rapidly than the nonrelativistic (cold dark and baryonic) matter density in a spatially flat universe. This eventually results in the expansion reaching a point at which the densities of nonrelativistic matter and DE have the same value and the decelerating cosmological expansion of the matter-dominated epoch switches to the accelerating expansion of the DE-dominated epoch that is currently observed \cite{MultRefs_4}.

In spacetime coordinates $x^\mu$ ($\mu = 0, 1, 2, 3$), with units chosen so that $\hbar = c = 1$, the late-time action of the model we consider is
\begin{eqnarray}
\label{action}
S&=&\int d^4x \ \sqrt{-g}\left[ \frac{m_p^{\phantom{p}2}}{16\pi}\left(-R+\frac{1}{2}g^{\mu\nu}\partial_\mu \Phi \partial_\nu \Phi \right.\right.\nonumber\\
&&\phantom{\int_\mathcal{D}d^4x \ \sqrt{-g}}\left.\phantom{\frac{m_p^{\phantom{p}2}}{16\pi}}\left.\phantom{\frac{1}{2}}-\frac{\kappa}{2} m_p^{\phantom{p}2}\Phi^{-\alpha} \right)+\mathcal{L}\right].
\end{eqnarray}
Here the Planck mass $m_p = G^{-1/2}$ where $G$ is the gravitational constant and $\mathcal{L}$ is the Lagrangian density of ordinary matter. The constants $\kappa$ and $\alpha$ are positive real numbers and we adopt
\begin{eqnarray}
\label{kappa}
\kappa = \dfrac{8}{3}\left( \dfrac{\alpha + 4}{\alpha + 2} \right) \left[ \dfrac{2}{3}\alpha(\alpha + 2) \right]^{\alpha /2}.
\end{eqnarray}
With this choice for $\kappa$, our results in the limit of zero space curvature reduce to those of Ref.\ \cite{Peebles&Ratra}.

Applying the variational principle with respect to the metric to the action (\ref{action}) gives the Einstein equations,
\begin{eqnarray}
\label{Einstein}
R_{\mu\nu} - \frac{1}{2}R g_{\mu\nu}=\frac{8\pi}{m_p^{\phantom{p}2}} &\left( T_{\mu\nu} +
Q_{\mu\nu} \right).
\end{eqnarray}
Here $R_{\mu\nu}$ and $R$ are the Ricci tensor and scalar and $T_{\mu\nu}$ is the stress-energy tensor of ordinary matter while $Q_{\mu\nu}$ is the stress-energy tensor of the $\Phi$ field and has the form
\begin{eqnarray}
\label{Q_tensor}
Q_{\mu\nu}&=&\frac{m_p^{\phantom{p}2}}{32\pi}\left[2\partial_\mu\Phi \partial_\nu\Phi -\left(g^{\zeta\xi} \partial_\zeta\Phi \partial_\xi\Phi  -\kappa\Phi^{-\alpha}\right)g_{\mu\nu}\right]. \nonumber\\
\end{eqnarray}

Assuming the cosmological principle of large-scale spatial homogeneity, the Friedmann metrics in coordinates $(t, r, \theta, \varphi)$ are
\begin{eqnarray}
\label{Fmetric}
ds^2 = dt^2 - a^2\left(\frac{dr^2}{1-k r^2}  + r^2d\theta^2  +r^2\sin^2\theta d\varphi^2\right).
\end{eqnarray}
Here $a$ is the scale factor and $k$ is the curvature parameter that takes values $-1, 0, 1$ for open, flat, and closed spatial geometry. References \cite{Peebles&Ratra,Ratra&PeeblesPRD} consider only the $k = 0$ case.

The equation of motion for the scale factor $a$ can be obtained by substituting the metric of Eq.\ (\ref{Fmetric}) into the Einstein equations (\ref{Einstein}). The equation of motion for the scalar field $\Phi$ can be derived by either applying the variational principle with respect to the $\Phi$ field to the scalar field part of the action (\ref{action}), or from the continuity conditions on the scalar field stress-energy tensor $Q_{\mu\nu}$ given in Eq. (\ref{Q_tensor}), and then using the metrics of Eq.\ (\ref{Fmetric}).

The complete system of equations of motion is
\begin{eqnarray}
\label{setofequations}
\ddot\Phi + 3\frac{\dot a}{a} \dot\Phi -\frac{\kappa\alpha}{2}m_p^{\phantom{p}2} \Phi^{-(\alpha+1)}&=&0,\\
\left(\frac{\dot a}{a}\right)^2&=&\frac{8\pi}{3m_p^{\phantom{p}2}}(\rho + \rho_\Phi) -\frac{k}{a^2}, \nonumber\\
\rho_\Phi&=&\frac{m_p^{\phantom{p}2}}{32\pi}\left(\dot\Phi^2 + \kappa m_p^{\phantom{p}2}\Phi^{-\alpha}\right). \nonumber
\end{eqnarray}
Here an overdot denotes a derivative with respect to time, and $\rho$ is the energy density of ordinary matter while $\rho_\Phi$ is that of the dark energy scalar field $\Phi$. It is also useful to introduce the density of spatial curvature,
\begin{eqnarray}
\label{rho_k}
\rho_k = -\frac{3m_p^{\phantom{p}2}}{8\pi}\dfrac{k}{a^2}.
\end{eqnarray}
In this convention the spatially open model has $\rho_k > 0$.

DE cannot have a significant effect at early times, so we assume $\rho_\Phi \ll \rho$ at $a(t) \ll a_0$, where $a_0$ is the current value of the scale factor. Neither can space curvature play a significant role in the early nonrelativistic matter-dominated epoch, so $\rho_k \ll \rho$ for $a(t) \ll a_0$. Under these assumptions the Einstein--de Sitter model provides an accurate description of the nonrelativistic matter-dominated epoch and so can be used to derive initial conditions for the scalar field $\Phi$ identical to these in the original flat-space case of Ref.\ \cite{Peebles&Ratra}. Of course, since the solution is a time-dependent fixed point or attractor, as shown here and in Refs.\ \cite{Peebles&Ratra, Ratra&PeeblesPRD}, it is not sensitive to the precise initial conditions adopted: a large range of initial conditions results in the same scalar field fixed point or attractor solution.

\subsection{Solution for the curvature-dominated epoch}
\label{CurvDomin}

In order to find whether the system (\ref{setofequations}) has an attractor solution in a certain epoch (i.e. matter dominated, radiation dominated or curvature dominated) we use a perturbation theory approach in which we treat the energy density of the scalar field $\Phi$ as a perturbation. Therefore, we neglect all terms in the right-hand side of the second equation of the system (\ref{setofequations}) (i.e. the Friedmann equation) except the energy density which dominates at the epoch of interest. When the energy budget of the Universe is dominated by radiation, ordinary matter or curvature, the solution of the Friedmann equation for the scale factor $a$ varies as a power of time, $a\propto t^n$ (which in general is not true in a quintessence-dominated epoch), where the index $n$ is determined for each epoch (as discussed in more detail later in this section). By substituting this power-law solution for the scale factor $a\propto t^n$ into the system (\ref{setofequations}), the equation of motion for the scalar field is
\begin{eqnarray}
\label{scalar field power law scale factor}
\ddot \Phi + \frac{3n}{t}\dot \Phi - \frac{\kappa\alpha}{2}m_{p}^{\phantom{p}2}\Phi^{-(\alpha + 1)} = 0.
\end{eqnarray}
Equation (\ref{scalar field power law scale factor}) has a special power-law solution 
\begin{eqnarray}
\label{Phi_e}
\Phi_e(t) = At^{2/(\alpha + 2)}
\end{eqnarray}
where the label $e$ denotes that this is an unperturbed, exact, spatially homogeneous solution. The value of the constant $A$ is
\begin{eqnarray}
\label{A_in_phi_e}
A&=&\left(\frac{\kappa\alpha m_{p}^{\phantom{p}2}(\alpha + 2)^2}{4[3n(\alpha + 2) -\alpha]}\right)^{1/(\alpha + 2)}.
\end{eqnarray}

We now show that, for the range of $\alpha$ and $n$ values that we are interested in, the special solution (\ref{Phi_e}) is an inwardly spiraling attractor in the phase space of solutions to (\ref{scalar field power law scale factor}).  This means, for example, that in a curvature-dominated epoch (which has $n = 1$), the scalar field will approach the special solution (\ref{Phi_e}) for a wide  range of initial conditions. 
In order to show this we follow the methods of Sec.\ V of Ref.\ \cite{Ratra&PeeblesPRD}, and make the  change of variables $(\Phi, t)\mapsto (u, \tau)$ where
\begin{eqnarray}
\label{Change_var}
\Phi(t) &=&\Phi_e(t)u(t), ~~~~ t=e^\tau.
\end{eqnarray}
Substituting (\ref{Change_var}) into (\ref{scalar field power law scale factor}) and using (\ref{Phi_e}) for
$\Phi_e(t)$ we derive the equation for perturbation $u(t)$ of the scalar field $\Phi(t)$,
\begin{eqnarray}
\label{pre_phasespace}
u''&-&\left(1-3n-\frac{4}{\alpha + 2}\right)u'\nonumber \\
&~&+\left(\frac{6n(\alpha + 2) - 2\alpha}{(\alpha + 2)^2}\right)\left(u - u^{-(\alpha + 1)}\right) = 0.
\end{eqnarray}
Here primes denote derivatives with respect to $\tau$. Finally we switch to the phase space of solutions of the system (\ref{scalar field power law scale factor}) by rewriting (\ref{pre_phasespace}) as the system
\begin{eqnarray}
\label{phasespace}
u'&=&p,\nonumber \\
p'&=&\left(1-3n-\frac{4}{\alpha + 2}\right)p\nonumber \\
&~&-\left(\frac{6n(\alpha + 2) - 2\alpha}{(\alpha + 2)^2}\right)\left(u - u^{-(\alpha + 1)}\right).
\end{eqnarray}
The critical point $(u_0, p_0) = (1, 0)$ corresponds to the special solution (\ref{Phi_e}). Although there  exist, in general,  other critical points at $p = 0$, these involve   complex roots of unity for $u$, which are not physically relevant in this case.

Taking the linearization of (\ref{phasespace}) about the critical point, one obtains the eigenvalues
\begin{eqnarray}
\label{lambda_12}
\lambda_{1,2} &=&f(\alpha,n) \pm i \sqrt{g(\alpha,n)}
\end{eqnarray}
where
\begin{eqnarray}
f(\alpha,n)&=&\frac{\alpha - 2 - 3n(\alpha + 2)}{2(\alpha + 2)},\\
g(\alpha,n)&=&\frac{6n(\alpha+2)(5\alpha+6)-9n^2(\alpha + 2)^2-(3\alpha + 2)^2}{4(2 + \alpha)^2}.\nonumber
\end{eqnarray}
For $f(\alpha, n)<0$ and $g(\alpha, n)>0$, the eigenvalues  $\lambda_1$ and $\lambda_2$ show that the critical point is an inwardly spiraling attractor in the phase space. The cases $n = 1/2$ (radiation-dominated epoch) and $n = 2/3$ (matter-dominated epoch) were previously  studied in Ref.\ \cite{Ratra&PeeblesPRD}. Note that for  the case $n = 1$ (curvature-dominated epoch) our critical point is  an inwardly spiraling attractor if $\alpha > -2 + 2/\sqrt{3}$ or if $\alpha <-4$. In the $\Phi$CDM model we are specifically  interested in the range $\alpha > 0$. So the critical point is an attractor for all $\alpha$ values of interest.

The above analysis ignores spatial inhomogeneities in the gravitational field. In the Appendix we show that the time-dependent fixed point solution found above remains stable in the presence of gravitational field inhomogeneities.

We can use our results to show how this model partially resolves the ``coincidence" puzzle. From the last equation of the system (\ref{setofequations}) it follows that in the curvature-dominated epoch
\begin{eqnarray}
\label{rho_Phi_t}
\rho_{\Phi}(t)&\propto & t^{-2\alpha /(\alpha + 2)},
\end{eqnarray}
while $\rho_k(t) \propto 1/t^2$ and $\rho_m(t) \propto 1/t^3$. The exponent in Eq.\ (\ref{rho_Phi_t}) varies from $-2$ to $0$ as $\alpha$ varies from $\infty$ to $0$, thus for $\alpha < \infty$ $\rho_\Phi(t)$ decays at a slower rate than $\rho_k$ in the curvature-dominated epoch and eventually comes to dominate. This is consistent with the results of similar analyses in the radiation-dominated and matter-dominated epochs given in Ref.\ \cite{Ratra&PeeblesPRD}.

\section{Some observational predictions}
\label{CosTests}
To gain some insight into the effects space curvature has on the $\Phi$CDM model, we compute predictions for some cosmological tests in this section. To make these predictions we first numerically integrate the equations of motion (\ref{setofequations}) with initial condition of the form (\ref{Phi_e}) taken in the matter-dominated epoch, where $n = 2/3$ with the usual expression for the scale factor in the matter-dominated epoch, see Ref.\ \cite{Peebles&Ratra}. Instead of $\rho$, $\rho_{\Phi}$ and $\rho_k$ we use dimensionless density parameters such as
\begin{eqnarray}
\label{Om}
\Omega_m = \dfrac{8\pi\rho}{3m_p^{\phantom{p}2}H^2} = \dfrac{\rho}{\rho + \rho_k + \rho_\Phi},
\end{eqnarray}
where $H = \dot{a}/a$ is the Hubble parameter. We present the predictions as isocontours in the space of model parameters ($\Omega_{m0}$,$\alpha$) for a number of different values of the spatial curvature density parameter $\Omega_{k0}$. (Here the subscript $0$ refers to the value at the current epoch. For the open model $\Omega_{k0} > 0$.) For our illustrative purposes here we consider the same four cosmological tests studied in Ref.\ \cite{Peebles&Ratra}. For a discussion of these and other cosmological tests see Ref.\ \cite{PeeblesRatra2003}. While it is of great interest to determine constrains on the three cosmological parameters of the model --- $\Omega_{m0}$, $\Omega_{k0}$, and $\alpha$ --- using various cosmological observables, in this paper we restrict ourselves to some qualitative remarks; a detailed quantitative comparison between the predictions of the model and observations is given in Ref.\ \cite{FarooqManiaRatra}, where it is found that observational data less tightly constrains space curvature in a dynamical dark energy matter of the type we study here.
\subsection{The time parameter $H_0t_0$}
\label{Ht_0}

\begin{figure}
        \includegraphics[width=0.5\textwidth]{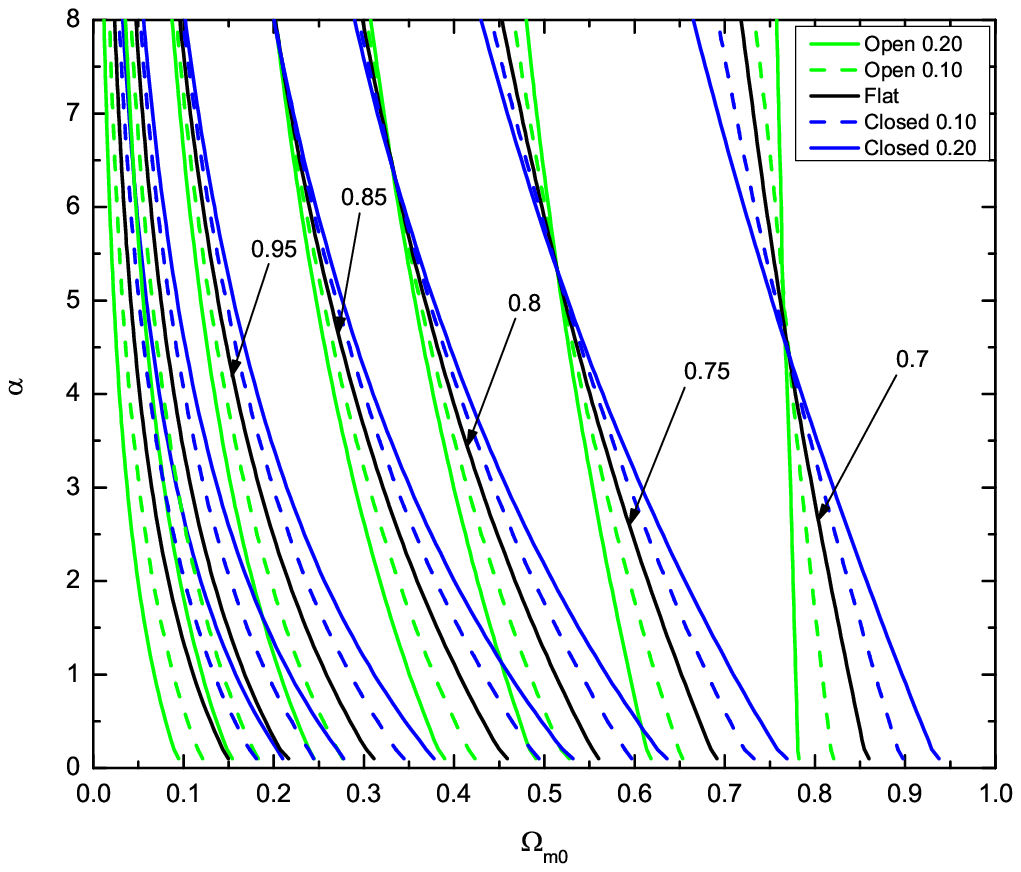}
        \includegraphics[width=0.5\textwidth]{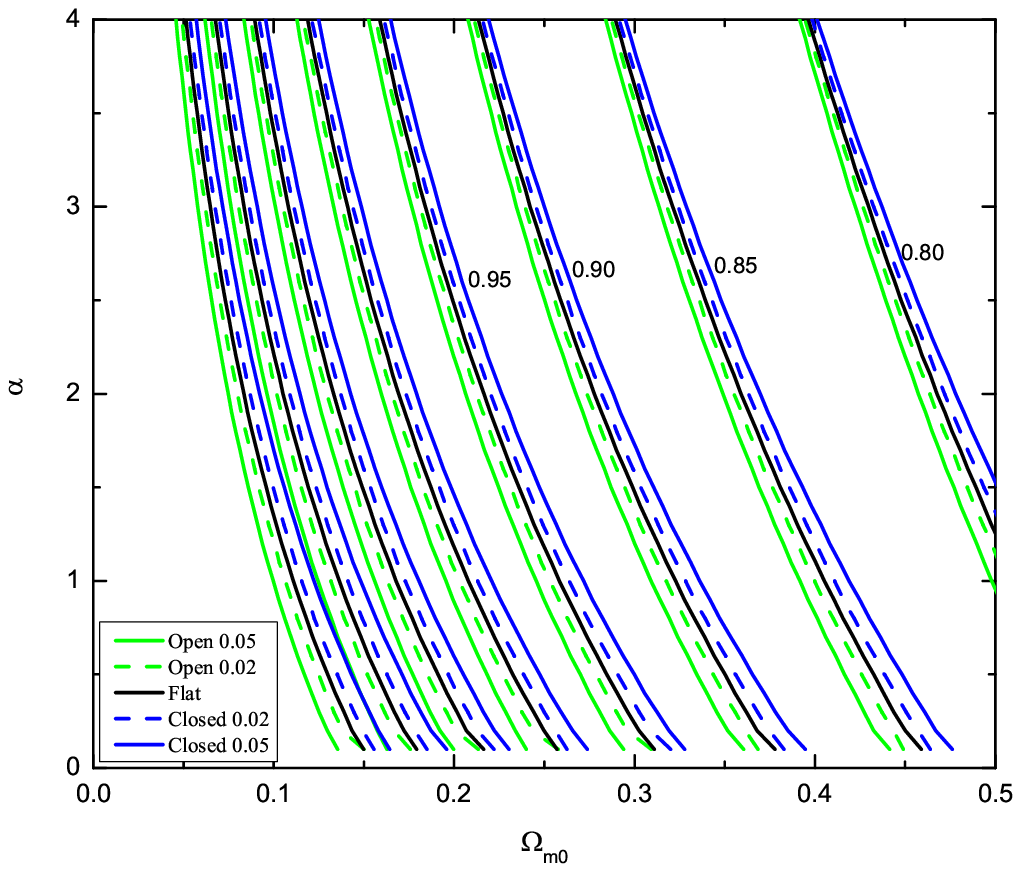}
        \caption{Contours of fixed time parameter $H_0t_0$, as a function of the present value of the nonrelativistic matter density parameter $\Omega_{m0}$ and scalar field potential power-law index $\alpha$, at various values of the current value of the space-curvature density parameter $\Omega_{k0}$ (as listed in the inset legend boxes). The upper panel shows a larger part of ($\Omega_{m0}$, $\alpha$) space for a larger range of $\Omega_{k0}$ values [for $H_0t_0 = 0.7, 0.75, 0.8, 0.85, 0.95, 1.05$ and $1.15$, from right to left], while the lower panel focuses on a smaller range of the three parameters [for $H_0t_0$ from $0.8$ to $1.15$ in steps of $0.05$, from right to left].}
        \label{fig: 1}
\end{figure}

The dimensionless time parameter is
\begin{eqnarray}
\label{H_0t_0Eq}
H_0t_0 = H_0\int_{0}^{a_0}\dfrac{da}{\dot{a}(t)},
\end{eqnarray}
where $t_0$ is the age of the Universe and $H_0$ and $a_0$ are the present values of the Hubble parameter and scale factor. Figure \ref{fig: 1} shows contours of constant $H_0t_0$ as a function of $\Omega_{m0}$ and $\alpha$ for a series of fixed values of $\Omega_{k0}$. A recent summary estimate of $H_0 = 68 \pm 2.8~{\rm km}~{\rm s}^{-1} {\rm Mpc}^{-1}$ \cite{ChenRatra2011} and the Planck (with WMAP polarization) estimate of $t_0 = 13.824 \pm~_{0.055}^{0.041}~{\rm Gyr}$ \cite{Ade} gives, for the $2\sigma$ range, $0.88 \leq H_0t_0 \leq 1.04$, where we have added the $1\sigma$ errors in quadrature and doubled to get the $2\sigma$ range. From Fig.\ \ref{fig: 1} we see that $\Omega_{m0} = 0.27$ and $\alpha = 3$ is reasonably consistent with these constraints for a range of $\Omega_{k0}$.

In the limit $\alpha \rightarrow 0$ this model reduces to the constant $\Lambda$ one (but not necessarily with zero space curvature), while the limit $\alpha \rightarrow \infty$ corresponds to the open, closed, or flat (Einstein--de Sitter) model with $\Lambda = 0$, depending on the value of space curvature. At fixed $\Omega_{m0}$ (and $\Omega_{k0}$), or in the flat-space case \cite{Peebles&Ratra}, the effect of increasing $\alpha$ is to reduce the value of $H_0t_0$, making the Universe younger at fixed $H_0$, since $\alpha = 0$ corresponds to a constant $\Lambda$ and so the oldest Universe for given $\Omega_{m0}$ and $\Omega_{k0}$. However, nonzero space curvature brings interesting new effects. At $\alpha = 0$ the $\Phi$CDM model reduces to the $\Lambda$CDM one and here it is well known that to hold $H_0t_0$ constant in the open case as $\Omega_{m0}$ is reduced and $\Omega_{k0}$ is increased requires a decrease in $\Omega_\Lambda$ (to compensate for the increase of $t_0$ at constant $H_0$ as $\Omega_{m0}$ is reduced and $\Omega_{k0}$ is increased). The converse is true in the closed case. Studying the $\alpha = 0$ intercepts of the $H_0t_0$ isocontours in both panels of Fig.\ \ref{fig: 1} confirms these arguments. That is, for a fixed value of $H_0t_0$ at smaller $\alpha$ (i.e. $\alpha \lesssim 4$) the contours corresponding to open geometry shift to the left of the flat geometry, i.e., to lower $\Omega_{m0}$, while the contours corresponding to closed geometry shift to the right of the flat case.

At higher $\alpha$ the DE density deceases more rapidly with the expansion (unlike the $\alpha = 0$ case where $\Lambda$ remains constant), and the contours switch around. Here to hold $H_0t_0$ constant in the open case as $\Omega_{k0}$ is increased requires that $\Omega_{m0}$ increase and $\Omega_{\Phi 0}$ decrease to compensate. In the closed case as $\Omega_{k0}$ is increased, $\Omega_{m0}$ must decrease and $\Omega_{\Phi 0}$ must also decrease. Thus, as evident from Fig.\ \ref{fig: 1}, for a given $H_0t_0$ value there is a point in ($\Omega_{m0}$, $\alpha$) space at which contours corresponding to different space curvatures cross. The intersection point moves to larger $\alpha$ as $\Omega_{m0}$ is decreased. This is because the Universe is older (at fixed $H_0$) at smaller $\Omega_{m0}$ so even DE with larger $\alpha$ now has more time to come to dominate the energy budget (and so behave more like DE with a constant DE density).

\subsection{The distance modulus difference $\Delta m(z)$}
\label{Delta_m}

\begin{figure}
   \includegraphics[width=0.5\textwidth]{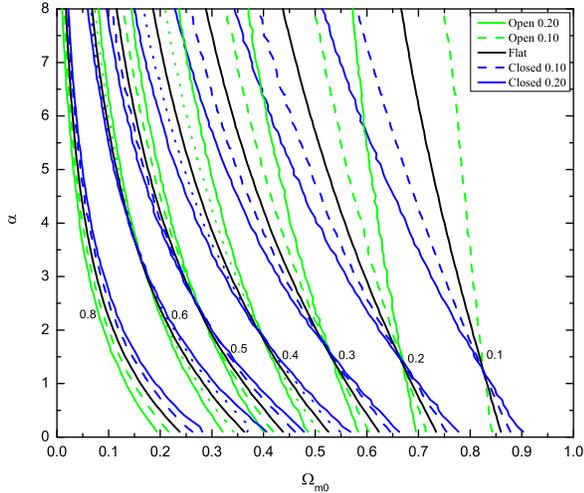}
   \includegraphics[width=0.5\textwidth]{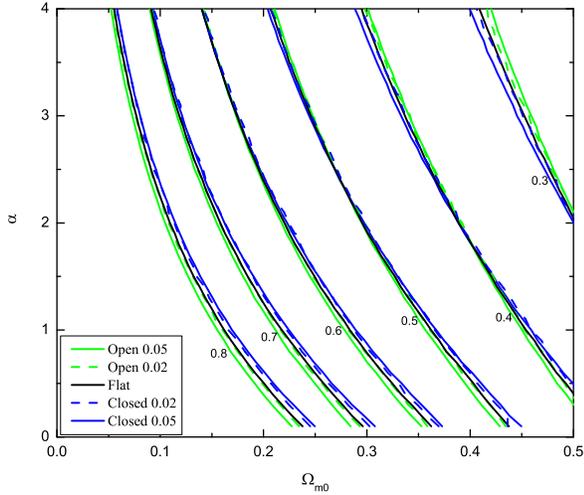}
    \caption{Contours of fixed bolometric distance modulus relative to the Einstein--de Sitter model, $\Delta m(z=1.5)$, as a function of the matter density parameter $\Omega_{m0}$ and scalar field potential power-law index $\alpha$, and various values of the space curvature density parameter $\Omega_{k0}$ (as listed in the inset legend boxes). The upper panel shows a larger part of ($\Omega_{m0}$, $\alpha$) space for a larger range of $\Omega_{k0}$ values [for $\Delta m(z=1.5) = 0.1, 0.2, 0.3, 0.4, 0.5, 0.6,$ and $0.8$ from right to left], while the lower panel focuses on a smaller range of the three parameters [for $\Delta m(z=1.5)$ from $0.3$ to $0.8$ in steps of $0.1$, from right to left]. In the upper panel there is no $\Omega_{k0} = 0.2$ contour for $\Delta m(z=1.5) = 0.1$ since in this case the model is too open for such a small distance modulus difference.}
 \label{fig: 2}
\end{figure}

We next consider the difference in bolometric distance moduli, at redshift $z = 1.5$, of the $\Phi$CDM model and the Einstein--de Sitter model. The coordinate distance $r$ is
\begin{eqnarray}
\label{coor.distance}
r = \dfrac{1}{\sqrt{-\Omega_{k0}}}\sin\left( \sqrt{-\Omega_{k0}}\int_{t_{\rm em}}^{t_0}\dfrac{dt}{a(t)} \right).
\end{eqnarray}
Here $t_{\rm em}$ and $t_0$ are the times when the signal was emitted and received.
Thus the difference in the distance moduli of the two models is
\begin{eqnarray}
\label{delta_mz}
\Delta m(z) = 5\log_{10}\left( \dfrac{r}{r_{\rm EdS}} \right)
\end{eqnarray}
where $r_{\rm EdS}$ is the coordinate distance in the Einstein--de Sitter model.

Figure \ref{fig: 2} shows contours of constant $\Delta m(z=1.5)$ as a function of $\Omega_{m0}$ and $\alpha$ for some values of $\Omega_{k0}$. Comparing Figs.\ \ref{fig: 1} and \ref{fig: 2}, we see that near $\alpha = 0$, where the DE behaves like constant $\Lambda$, $\Delta m(z=1.5)$ is less sensitive to the value of $\Omega_{k0}$ than is $H_0t_0$. However at larger $\alpha$ $\Delta m(z=1.5)$ is more sensitive to spatial curvature  than is $H_0t_0$. Clearly, extending the $\Phi$CDM model to include space curvature as a free parameter broadens the range of allowed parameter values. As in the $H_0t_0$ case, for a given value of $\Delta m(z=1.5)$ there is a point in ($\Omega_{m0}$, $\alpha$) space at which all contours intersect.

\subsection{Number counts}
\label{NumCounts}

\begin{figure}
   \includegraphics[width=0.5\textwidth]{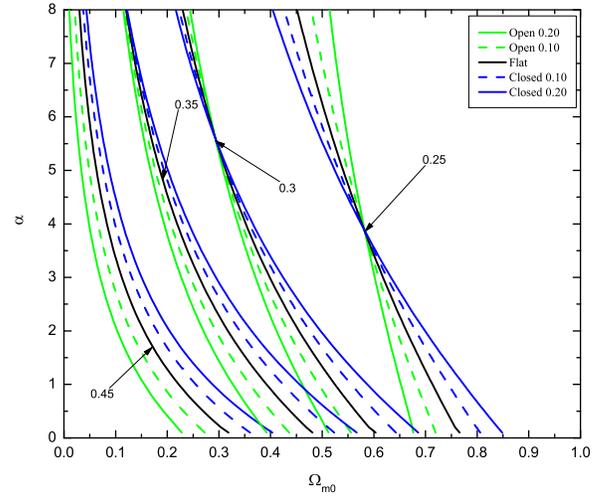}
   \includegraphics[width=0.5\textwidth]{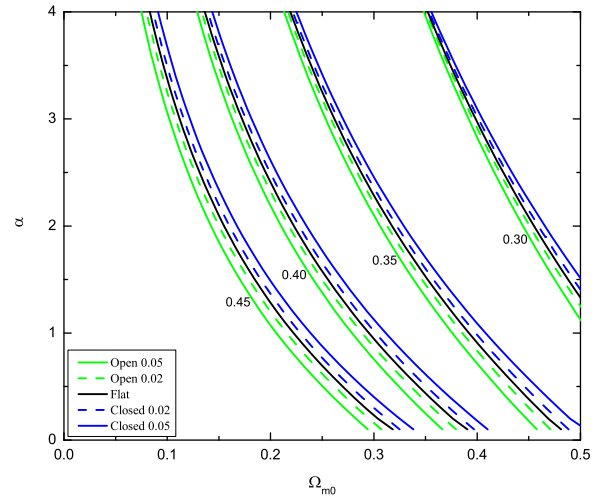}
    \caption{Contours of fixed $A(z=0.7)$ as a function of $\Omega_{m0}$ and $\alpha$ at various values of $\Omega_{k0}$ (as listed in the inset legend boxes). The upper panel shows a larger part of the parameter space for $A(z=0.7) = 0.25, 0.3, 0.35$ and $0.45$ from right to left. The lower panel shows a smaller range of the three parameters for $A(z=0.7) = 0.3, 0.35, 0.4$ and $0.45$ from right to left.}
 \label{fig: 3}
\end{figure}

The count per unit increment of redshift for conserved objects is\footnote{See Sec. IV.B.5 of Ref. \cite{PeeblesRatra2003} and Refs. \cite{MultRefs_add1} for discussions of this test.}
\begin{eqnarray}
\label{Az}
\dfrac{dN}{dz} \propto z^2A(z), ~A(z) = \dfrac{H_0^{\phantom{0}3}a_0^{\phantom{0}2}r^2}{z^2}\dfrac{a}{\dot{a}}.
\end{eqnarray}
Isocontours of fixed $A(z=0.7)$ are shown in Fig.\ \ref{fig: 3}. The general features are similar to those shown in Figs.\ \ref{fig: 1} and \ref{fig: 2} for $H_0t_0$ and $\Delta m(z=1.5)$.

\subsection{The growth of structure}
\label{growth}

\begin{figure*}
\centering
  \begin{tabular}{@{}cc@{}}
    \includegraphics[width=0.5\textwidth]{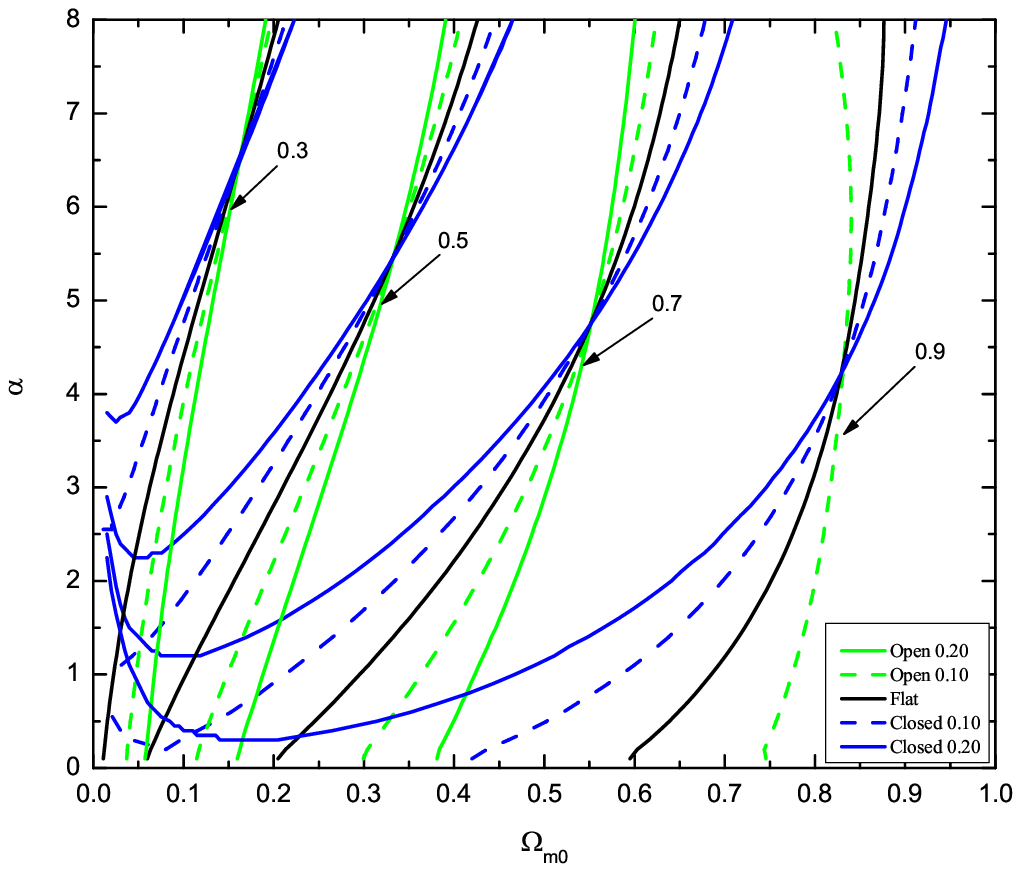}&
    \includegraphics[width=0.5\textwidth]{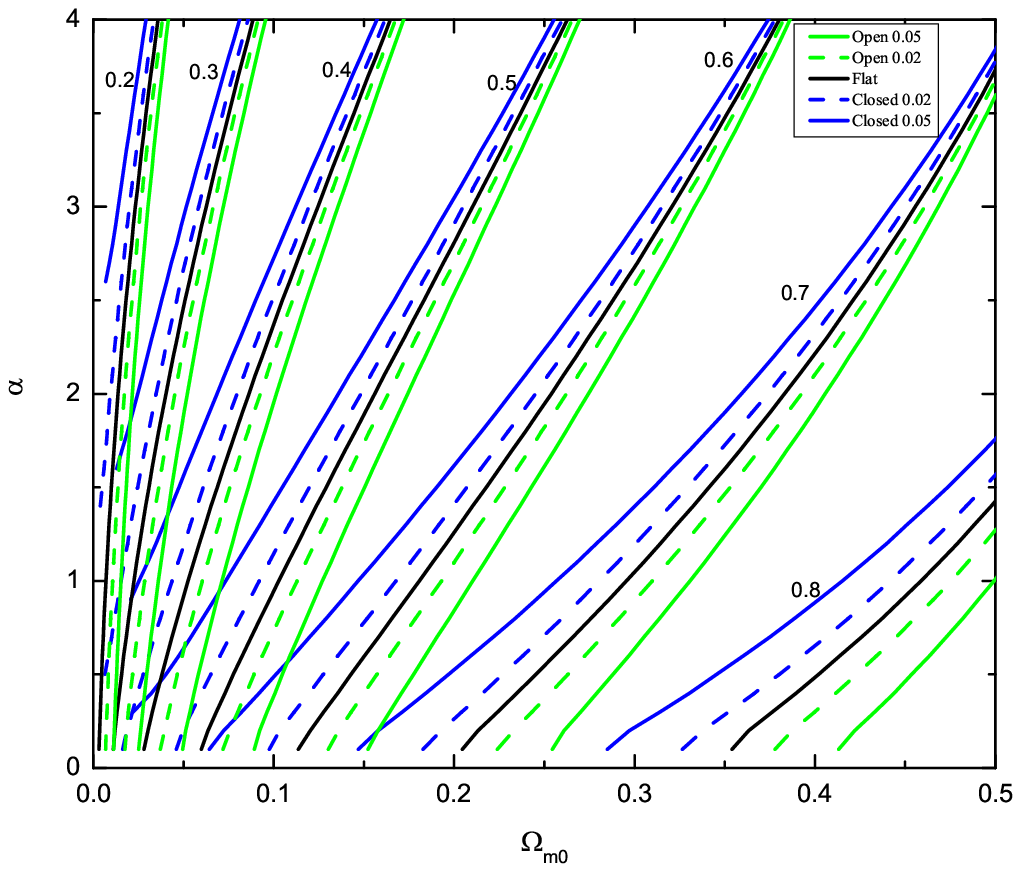} \\
  \end{tabular}
    \caption{Contours of the factor by which the growth of ordinary matter perturbations falls below that of the Einstein--de Sitter model, $\Delta(\Omega_{m0}, \Omega_{k0}, \alpha)$, as a function of the matter density parameter $\Omega_{m0}$ and scalar field potential power-law index $\alpha$, and various values of the space curvature density parameter $\Omega_{k0}$ (as listed in the inset legend boxes). The left panel shows a larger part of ($\Omega_{m0}$, $\alpha$) space for a larger range of $\Omega_{k0}$ values [for $\Delta(\Omega_{m0}, \Omega_{k0}, \alpha) = 0.3, 0.5, 0.7$ and $0.9$ from left to right], while the right panel focuses on a smaller range of the three parameters [for $\Delta(\Omega_{m0}, \Omega_{k0}, \alpha)$ from $0.2$ to $0.8$ in steps of $0.1$, from left to right]. In the left panel there is no $\Omega_{k0} = 0.2$ contour for $\Delta = 0.9$ since in this case the model is too open to allow such a large growth factor.}
 \label{fig: 4}
\end{figure*}

Finally, we consider the growth of large-scale structure of the Universe which started as small primordial density inhomogeneities in the early Universe \cite{MultRefs_5}. Within the framework of linear perturbation theory the scalar filed stays homogeneous as we show in the Appendix on the scales of matter perturbations and the density contrast in ordinary matter, $\delta = \delta \rho/\rho$, satisfies
\begin{eqnarray}
\label{gr_factor}
\ddot{\delta} + 2\dfrac{\dot{a}}{a}\dot{\delta} - \dfrac{4\pi}{m_p^{\phantom{p}2}}\rho\delta = 0.
\end{eqnarray}
Following Ref.\ \cite{Peebles&Ratra} the cosmological test parameter we consider is
\begin{eqnarray}
\label{DELTA}
\Delta(\Omega_{m0}, \Omega_{k0}, \alpha) = \dfrac{\delta(t_0)}{(1 + z_i)\delta(t_i)}
\end{eqnarray}
where $t_0$ denotes the current epoch while $t_i$ is the time when the scale factor $a_i \ll a_0$, well within the matter-dominated epoch when the Einstein--de Sitter model was a good approximation. Thus the factor $\Delta$ is the ratio by which the growth of linear fluctuations in density have declined below that of the Einstein--de Sitter model prediction. We graph contours of $\Delta$ in Fig.\ \ref{fig: 4}.

There are two interesting facts about the $\Delta$ contours shown in Fig.\ \ref{fig: 4}. First, the growth rate is quite sensitive to the value of $\Omega_{k0}$, much more so than any of the other parameters we have considered. (This is not unexpected, as it is well known in more conventional models that the growth factor is much more sensitive to $\Omega_{m0}$ when $\Omega_{k0}$ is nonzero.) Second, the curvature dependence of the isocontours is the opposite of that for the other three parameters. So a joint analysis of growth factor and geometry measurements would seem to be a very good way to constrain $\Omega_{k0}$.

\section{Conclusion}
\label{concl}

We have extended the $\Phi$CDM model to nonflat geometries and shown that in the curvature-dominated epoch the solution is also an attractor or time-dependent fixed point (see Sec.\ \ref{CurvDomin} and the Appendix). We have computed predictions of the model for an illustrative set of cosmological tests and shown that the presence of space curvature will broaden the allowed range of model parameters. Spatial curvature should be considered as a free parameter when observational data are analyzed. The nonflat $\Phi$CDM model we have developed here is the only consistent nonflat time-variable DE model to date and can be used as a fiducial model for such analyses.

Our computations have shown that for a single cosmological test there is a degeneracy point in parameter space for each fixed value of the cosmological observable of the test. At this point one cannot differentiate between contours corresponding to different values of spatial curvature. However, these points of degeneracy do not coincide in ($\Omega_{m0}$, $\alpha$) parameter space for the different cosmological tests. Hence it is important to use multiple cosmological tests in order to determine spatial curvature from observations.

We have noted that a joint analysis of geometry and growth factor measurements appears to be a fruitful way to constrain space curvature. CMB anisotropy data will also likely provide useful constraints on space curvature. This will first require accounting for spatial curvature effects on the quantum-mechanical zero-point fluctuations generated during inflation, which will affect the primordial density perturbations power spectrum \cite{MultRefs_6}. While the curved-space computation is more involved than the corresponding flat-space one, the resulting constraints from CMB anisotropy data on space curvature in the presence of dynamical dark energy are likely to prove quite interesting.

\section{Acknowledgments}
\label{Acknegt}

We thank our colleagues Omer Farooq and Data Mania for valuable suggestions and for checking our computations. This work was supported by DOE Grant No. DEFG03-99GP41093 and NSF Grant No. AST-1109275.
\appendix
\section{}
\label{appx}

In Sec.\ \ref{CurvDomin} we showed that the special time-dependent fixed point solution for the scalar field in the curvature-dominated epoch is stable if we ignore spatial inhomogeneities in the gravitational field. In this Appendix we show that gravitational spatial inhomogeneities do not spoil this property of the solution, thus preserving the inclination of the scalar field DE density to always want to try to dominate over the dominant energy density source \cite{Peebles&Ratra, Ratra&PeeblesPRD}.

Inhomogeneities in the scalar field will induce inhomogeneities in the metric, and vice versa. We show that, in the curvature-dominated epoch, any slight inhomogeneities will die out. (This generalizes the flat-space results of Sec.\ IX of Ref.\ \cite{Ratra&PeeblesPRD}.)

We linearize the disturbances in the metric about a curved Friedmann background metric in synchronous gauge. To this end, we write the line element as
\begin{eqnarray}
\label{A1}
ds^2 = \tilde g_{\mu\nu}dx^\mu dx^\nu = (g_{\mu\nu} + \delta g_{\mu\nu})dx^\mu dx^\nu.
\end{eqnarray}
We work in time-orthogonal coordinates $(t, r, \theta, \varphi)$ with $g_{\mu \nu}$ given in Eq.\ (\ref{Fmetric}) and the perturbations
\begin{eqnarray}
\delta g_{\mu\nu} = a(t)^2\left(\begin{array}{cccc}
0& 0 & 0 & 0\\
0 &f(r)h_{rr}&h_{r\theta} &h_{r\varphi}\\
0 &h_{r\theta} & r^2 h_{\theta\theta} &h_{\theta\varphi}\\
0 &h_{r\varphi} &h_{\theta\varphi} & r^2\sin^2(\theta) h_{\varphi\varphi}
\end{array}\right), \nonumber\\
\label{A2}
\end{eqnarray}
where $f(r) = 1/(1 - kr^2)$, $|h_{ij}| \ll 1$, and each $h_{ij}$ is a function of $t, r,\theta$, and $\varphi$.

The scalar field equation of motion in a space-time with cometric $\tilde{g}^{\mu\nu}$ reads
\begin{eqnarray}
\label{A3}
\tilde \nabla_\mu (\tilde{g}^{\mu\nu} \partial_\nu \Phi ) + V'(\Phi) = 0.
\end{eqnarray}
The perturbed scalar field is written
\begin{eqnarray}
\label{A4}
\Phi(x^\mu) = \Phi_0(t) + \phi(x^\mu)
\end{eqnarray}
where $\phi$ is a small perturbation, $|\phi| \ll |\Phi_0|$, and $\Phi_0$ is a solution to the scalar field equation of motion in the unperturbed homogeneous Friedmann background,
\begin{eqnarray}
\label{A5}
\ddot\Phi_0 + 3\frac{\dot a}{a}\dot\Phi_0 -\frac{\kappa \alpha m_p^2}{2}\Phi_0^{-(\alpha + 1)} = 0.
\end{eqnarray}
Plugging (\ref{A4}) into (\ref{A5}) gives, to first order in $\phi$,
\begin{eqnarray}
\label{A6}
\ddot\phi + \frac{3\dot a}{a}\dot\phi-\frac{1}{a^2}\nabla^2 \phi + V''(\Phi_0) \phi - \frac{1}{2}\dot h \dot\Phi_0 = 0,
\end{eqnarray}
\\where $h = h_{rr} + h_{\theta\theta} + h_{\varphi\varphi} = -g^{\mu\nu}\delta g_{\mu\nu}$, and $\nabla^2$ is the Laplacian for the three-dimensional spacelike hypersurface of constant $t$ in the unperturbed Friedman geometry,
\begin{eqnarray}
\label{A7}
\nabla^2 &=& \frac{1}{r^2}\frac{\partial}{\partial r}\left((r^2 - k r^4)\frac{\partial }{\partial r}\right) + k
r\frac{\partial }{\partial r} \\
&+&\frac{1}{r^2\sin(\theta)}\frac{\partial}{\partial \theta}\left(\sin(\theta) \frac{\partial }{\partial \theta}\right)
+\frac{1}{r^2\sin^2(\theta)}\frac{\partial^2 }{\partial \varphi^2}.\nonumber
\end{eqnarray}
When $k = 0$ $\nabla^2$ is the usual three-dimensional flat-space Laplacian in spherical coordinates. 

The $tt$ component of the stress-energy tensor $Q_{\mu\nu}$ for $\Phi = \Phi_0 + \phi$, to first order in $\phi$, is
\begin{eqnarray}
\label{A8}
Q_{tt} = \frac{m_p^2}{32\pi}\left[\dot\Phi_0^2+2V(\Phi_0)\right]\nonumber\\
+\frac{m_p^2}{16\pi}\left[\dot\Phi_0\dot\phi + V'(\Phi_0)\phi\right],
\end{eqnarray}
and the trace $Q = \tilde g^{\mu\nu}Q_{\mu\nu}$ is, to first order,
\begin{eqnarray}
\label{A9}
Q = \frac{m_p^2}{16\pi}\left[4V(\Phi_0) - \dot\Phi_0^2\right]\nonumber\\
+\frac{m_p^2}{8\pi}\left[2\phi V'(\Phi_0) - \dot\Phi_0\dot \phi \right].
\end{eqnarray}
As for the Ricci tensor $R_{\mu\nu}$, we will also only require the $tt$ component. To first order it is
\begin{eqnarray}
\label{A10}
R_{tt} = -\frac{3\ddot a}{a} +\left[\frac{\dot a}{a}\dot h + \frac{1}{2}\ddot h\right].
\end{eqnarray}
By the Einstein field equations (\ref{Einstein}) we therefore get the first-order perturbation equation,
\begin{eqnarray}
\label{A11}
\ddot h + \frac{2\dot a}{a}\dot h = 2\dot \Phi_0 \dot\phi -V'(\Phi_0)\phi.
\end{eqnarray}
This corresponds to Eq.\ (3.14) of Ref.\ \cite{Ratra&PeeblesPRD1995}.

We now take $a=a_0 t$ for the curvature-dominated epoch, where $a_0$ is a constant of integration and we consider times $t > 0$. Thus, the system we need to analyze is
\begin{eqnarray}
\label{A12}
\ddot\phi + \frac{3}{t}\dot\phi-\frac{L^2}{a_0^2 t^2} \phi + V''(\Phi_0) \phi = \frac{1}{2}\dot h \dot\Phi_0,
\end{eqnarray}
\begin{eqnarray}
\label{A13}
\ddot h + \frac{2}{t}\dot h = 2\dot \Phi_0 \dot\phi -V'(\Phi_0)\phi.
\end{eqnarray}
Here we have made a hyperspherical  harmonic transformation, the variables $\phi$ and $h$ are now harmonic mode amplitudes, and $L^2$ is the eigenvalue of the Laplacian operator (\ref{A8}) (see Ref.\ \cite{Ratra&PeeblesPRD1995} and Sec.\ II of Ref.\ \cite{RatraPRD1994}).  One has $L^2\rightarrow -1$ (respectively $L^2\rightarrow-\infty$) in the limit of long wavelength (short wavelength) modes for the negative curvature case.

The field $\Phi_0$ is the special solution obtained in Sec.\ \ref{CurvDomin}, Eq.\ (\ref{Phi_e}). We here write it as
\begin{eqnarray}
\label{A14}
\Phi_0 = At^m,
\end{eqnarray}
where
\begin{eqnarray}
\label{A15}
m = \frac{2}{\alpha + 2},
\end{eqnarray}
and  $A$ is, by (\ref{A_in_phi_e}) with $n = 1$,
\begin{eqnarray}
\label{A16}
A = \left(\frac{\kappa \alpha m_p^2(\alpha + 2)}{8\alpha + 24}\right)^{1/(\alpha + 2)}.
\end{eqnarray}
Defining
\begin{eqnarray}
B = \frac{\kappa\alpha}{2}m_p^2,
\label{A17}
\end{eqnarray}
Eqs.\ (\ref{A12}) and (\ref{A13}) can be rewritten as
\begin{eqnarray}
\label{A18}
&\ddot\phi &~+~\frac{3}{t}\dot\phi+\frac{J}{t^2}\phi = \frac{mA}{2}\dot h t^{m-1},\\
\nonumber \\
&\ddot h &~+~\frac{2}{t}\dot h = 2mAt^{m-1}\dot\phi +BA^{-(\alpha + 1)}t^{m-2}\phi,
\label{A19}
\end{eqnarray}
where $J=(\alpha + 1)(m^2 + 2m) - L^2/a_0^2$. As mentioned previously, $L^2 \rightarrow -1$ in the  case that we are presently interested in (long-wavelength perturbations and negative curvature), so the constant $J$ is a positive real number $> 3$.

For the curvature-dominated case $\rho_k \propto t^{-2}$ and so
\begin{eqnarray}
\label{A20}
C^2 \frac{\rho_\Phi}{\rho_k} = t^{2m},
\end{eqnarray}
where $C$ is a constant of integration. Thus, Eqs. (\ref{A18}) and (\ref{A19}) can be written as
\begin{eqnarray}
\label{A21}
\ddot\phi + \frac{3}{t}\dot\phi+\frac{J}{t^2}\phi = \frac{mAC}{2} \frac{\dot h}{t} \sqrt{\frac{\rho_\Phi}{\rho_k}},
\end{eqnarray}
\begin{eqnarray}
\label{A22}
\ddot h + \frac{2}{t}\dot h = \frac{2mBC}{t}\sqrt{\frac{\rho_\Phi}{\rho_k}}\dot\phi +\frac{BA^{-(\alpha + 1)}C}{t^{2}}\sqrt{\frac{\rho_\Phi}{\rho_k}}\phi. \nonumber\\
\end{eqnarray}
Following Ref.\ \cite{Ratra&PeeblesPRD} we solve these equations by using a linear perturbation technique. Since we are in the curvature-dominated epoch and $\sqrt{\rho_\Phi/\rho_k}$ is small, we begin by searching for approximate solutions to (\ref{A21}) and (\ref{A22}) where the source terms on the right-hand side are neglected. That is, we first solve the homogeneous equations (to get zeroth order solutions for  $\phi$ and $\dot h$),
\begin{eqnarray}
\label{A23}
\ddot\phi_0 + \frac{3}{t}\dot\phi_0+\frac{J}{t^2}\phi_0 = 0,
\end{eqnarray}
\begin{eqnarray}
\ddot h_0 + \frac{2}{t}\dot h_0 = 0,
\label{A24}
\end{eqnarray}
where subscript $0$ now denotes solutions in zeroth order of the perturbation approach. Once we have these zeroth order solutions, we will plug them into the right-hand side of Eqs. (\ref{A18}) and (\ref{A19}) in order to obtain new differential equations which can then be used to derive correction terms of order $\sqrt{\rho_\Phi/\rho_k}$. If our solutions with order $\sqrt{\rho_\Phi/\rho_k}$ corrections are still decaying then it means that the stability result is established at least in the first-order perturbation analysis.

The zeroth order solution to (\ref{A23}) is
\begin{eqnarray}
\label{A25}
\phi_0(t) = \dfrac{C_1 }{t}\cos\left[\sqrt{J-1} \ln (t)\right]\nonumber\\
+\dfrac{C_2 }{t}\sin\left[\sqrt{J-1}\ln (t)\right],
\end{eqnarray}
where $C_1$ and $C_2$ are constants of integration, and the zeroth order solution to (\ref{A24}) is
\begin{eqnarray}
\label{A26}
h_0(t) = \dfrac{C_3}{t} + C_4,
\end{eqnarray}\\
where $C_3$ and $C_4$ are constants of integration. Note that, up to oscillatory bounded functions of time, $\phi_0/\Phi_0 \propto t^{-(\alpha + 4)/(\alpha + 2)}~ \in (t^{-2}, t^{-1})$, so we confirm the result of Sec.\ \ref{CurvDomin} that if we ignore the effect of metric perturbations the time-dependent fixed-point solution is stable.

Writing $\phi = \phi_0 + \phi_1$ and $h = h_0 + h_1$, and plugging (\ref{A26}) into Eq. (\ref{A18}), we get for the first order $\phi_1$ equation
\begin{eqnarray}
\label{A27}
\ddot\phi_1 + \frac{3}{t}\dot\phi_1 + \frac{J}{t^2} \phi_1 = \frac{mA}{2}\dot h_0 t^{m-1}.
\end{eqnarray}
(We shall not need the $h_1$ differential equation.) Solving this differential equation for $\phi_1$ we find
\begin{eqnarray}
\label{A28}
\phi_1(t) = -\dfrac{mAC_3t^{m-1}}{2(m^2 - 1 + J)}.
\end{eqnarray}
From this solution and that in (\ref{A25}), we find, up to oscillatory bounded functions of time, $\phi_1(t)/\phi_0(t) \propto \sqrt{\rho_\Phi/\rho_k}$, so in the curvature-dominated epoch, where $\rho_\Phi \ll \rho_k$, the correction to the scalar field solution from the metric inhomogeneity is small.


\begin{thebibliography}{99}

\bibitem{Peebles}P. J. E. Peebles, Astrophys. J. {\bf 284}, 439 (1984).

\bibitem{MultRefs_1}For early indications, see e.g., H. K. Jassal, J. S. Bagla and T. Padmanabhan, Mon. Not. R. Astron. Soc. {\bf 405}, 2639 (2010); K. M. Wilson, G. Chen and B. Ratra, Mod. Phys. Lett. A {\bf 21}, 2197 (2006); T. M. Davis et al., Astrophys. J. {\bf 666}, 716 (2007); S. W. Allen et al., Mon. Not. R. Astron. Soc. {\bf 383}, 879 (2008).

\bibitem{Weinberg}D. H. Weinberg et al., arxiv:1306.0913.

\bibitem{Ade}See G. Hinshaw et al. (WMAP Collaboration), Astrophys. J. Suppl. {\bf 208}, 19 (2013); P. A. R. Ade et al. (Planck Collaboration), arxiv:1303.5076; for an early indication, see S. Podariu et al. Astrophys. J. {\bf 559}, 9 (2001).

\bibitem{MultRefs_2}Early work includes R. Aurich and F. Steiner, Mon. Not. R. Astron. Soc. {\bf 334}, 735 (2002); Int. J. Mod. Phys. D {\bf 13}, 123 (2004); Phys. Rev. D {\bf 67}, 123511 (2003); J. L. Crooks et al., Astropart. Phys. {\bf 20}, 361 (2003); K. Ichikawa and T. Takahashi,  Phys. Rev. D {\bf 73}, 083526 (2006); J. Cosmol. Astropart. Phys. {\bf 02}, (2007) 001; {\bf 04} (2008) 027; E. L. Wright, arxiv:astro-ph/0603750; K. Ichikawa et al., J. Cosmol. Astropart. {\bf 12}, (2006) 005; G.-B. Zhao et al., Phys. Lett. B {\bf 648}, 8 (2007); Y. Wang and P. Mukherjee, Phys. Rev. D {\bf 76}, 103533 (2007); Y. Gong, Q. Wu and A. Wang, Astrophys. J. {\bf 681}, 27 (2008); J.-M. Virey et al., J. Cosmol. Astropart. Phys. {\bf 12}, (2008) 008; M. J. Mortonson, Phys. Rev. D {\bf 80}, 123504 (2009).

\bibitem{RatraPRD1991}B. Ratra, Phys. Rev. D {\bf 43}, 3802 (1991); S. Podariu and B. Ratra, Astrophys. J. {\bf 532}, 109 (2000).

\bibitem{Peebles&Ratra}P. J. E. Peebles and B. Ratra, Astrophys. J. Lett. {\bf 325}, L17 (1988).

\bibitem{Ratra&PeeblesPRD}B. Ratra and P. J. E. Peebles, Phys. Rev. D {\bf 37}, 3406 (1988).

\bibitem{JMartin2008}J. Martin, Mod. Phys. Lett. A {\bf 23}, 1252 (2008).

\bibitem{MultRefs_3}R. Aurich and F. Steiner, Mon. Not. R. Astron. Soc. {\bf 334}, 735 (2002); Int. J. Mod. Phys. D {\bf 13}, 123 (2004); Phys. Rev. D {\bf 67}, 123511 (2003); K. Thepsuriya and B. Gumjudpai, arxiv:0904.2743; Z.-Q. Chen and D.-H. Guo, Int. J. Theor. Phys. {\bf 51}, 3856 (2012); B. Gumjudpai and K. Thepsuriya, Astrophys. Space Sci. {\bf 342}, 537 (2012).

\bibitem{MultRefs_4}N. G. Busca et al., Astron. Astrophys. {\bf 552}, A96 (2013); O. Farooq and B. Ratra, Astrophys. J. Lett. {\bf 766}, L7 (2013); O. Farooq, S. Crandall and B. Ratra, Phys. Lett. B {\bf 726}, 72 (2013).

\bibitem{PeeblesRatra2003}P. J. E. Peebles and B. Ratra, Rev. Mod. Phys. {\bf 75}, 559 (2003).

\bibitem{FarooqManiaRatra}O. Farooq, D. Mania and B. Ratra, arxiv:1308.0834.

\bibitem{ChenRatra2011}G. Chen and B. Ratra, Publ. Astron. Soc. Pac. {\bf 123}, 1127 (2011).

\bibitem{MultRefs_add1}J. A. Newman and M. Davis, Astrophys. J. Lett. {\bf 534}, L11 (2000); D. Huterer and M. S. Turner, Phys. Rev. D {\bf 64}, 123527 (2001); S. Podariu and B. Ratra, Astrophys. J. {\bf 563}, 28 (2001).

\bibitem{MultRefs_5}W. Fisher, B. Ratra and L. Susskind, Nucl. Phys. B{\bf 259}, 730 (1985); B. Ratra, Phys. Rev. D {\bf 45}, 1913 (1992), and references therein.

\bibitem{MultRefs_6}J. R. Gott, Nature (London) {\bf 295}, 304 (1982); B. Ratra and P. J. E. Peebles, Astrophys. J. Lett. {\bf 432}, L5 (1994); Phys. Rev. D {\bf 52}, 1837 (1995); M. Kamionkowski et al., Astrophys. J. {\bf 434}, L1 (1994); M. Bucher, A. S. Goldhaber and N. Turok, Phys. Rev. D {\bf 52}, 3314 (1995); D. H. Lyth and A. Woszczyna, Phys. Rev. D {\bf 52}, 3338 (1995); K. Yamamoto, M. Sasaki and T. Tanaka, Astrophys. J. {\bf 455}, 412 (1995); K. Ganga et al., Astrophys. J. {\bf 484}, 517 (1997); M. G{\'o}rski et al., Astrophys. J. Suppl. {\bf 114}, 1 (1998).

\bibitem{Ratra&PeeblesPRD1995}B. Ratra and P. J. E. Peebles, Phys. Rev. D {\bf 52}, 1837 (1995).

\bibitem{RatraPRD1994}B. Ratra, Phys. Rev. D {\bf 50}, 5252 (1994).

\end{thebibliography}
\end{document}